\documentclass[letterpaper]{article} 
\usepackage{aaai25}  
\usepackage{times}  
\usepackage{helvet}  
\usepackage{courier}  
\usepackage[hyphens]{url}  
\usepackage{graphicx} 
\usepackage{natbib}  
\usepackage{caption} 
\usepackage{algorithm}
\usepackage{algorithmic}

\usepackage{xcolor}

\usepackage{newfloat}
\usepackage{listings}
\DeclareCaptionStyle{ruled}{labelfont=normalfont,labelsep=colon,strut=off} 
\floatstyle{ruled}
\newfloat{listing}{tb}{lst}{}
\floatname{listing}{Listing}
\title{The Impact of Big Five Personality Traits on AI Agent Decision-Making in Public Spaces: A Social Simulation Study}
\author{  
    Minjun Ren\textsuperscript{\rm 1},
    Wentao Xu\textsuperscript{\rm 2}
    \thanks{Corresponding author, myrainbowandsky@gmail.com}
}
\affiliations{
    \textsuperscript{\rm 1} School of Computer Science and Technology,  \\ \textsuperscript{\rm 2} Department of Science and Technology of Communication \\ 
    
    \textsuperscript{\rm 1,2} University of Science and Technology of China\\
    No.96, JinZhai Street, Hefei, Anhui, 230026, China\\
}

\begin{document}

\maketitle

\begin{abstract}
This study investigates how the Big Five personality traits influence decision-making processes in AI agents within public spaces. Using AgentVerse framework and GPT-3.5-turbo, we simulated interactions among 10 AI agents, each embodying different dimensions of the Big Five personality traits, in a classroom environment responding to misinformation. The experiment assessed both public expressions ([Speak]) and private thoughts ([Think]) of agents, revealing significant correlations between personality traits and decision-making patterns. Results demonstrate that Openness to Experience had the strongest impact on information acceptance, with curious agents showing high acceptance rates and cautious agents displaying strong skepticism. Extraversion and Conscientiousness also showed notable influence on decision-making, while Neuroticism and Agreeableness exhibited more balanced responses. Additionally, we observed significant discrepancies between public expressions and private thoughts, particularly in agents with friendly and extroverted personalities, suggesting that social context influences decision-making behavior. Our findings contribute to understanding how personality traits shape AI agent behavior in social settings and have implications for developing more nuanced and context-aware AI systems.
\end{abstract}

%

\section{Introduction}
With the rapid advancement of artificial intelligence technology, AI agents have demonstrated increasingly sophisticated capabilities in simulating human behavior and decision generation. \citet{park2023generativeagentsinteractivesimulacra} explored in their study how generative agents achieve more natural interactions by simulating human behavior. Furthermore, \citet{sorokovikova2024llmssimulatebigpersonality} demonstrated that large language models (LLMs) can simulate the Big Five personality traits, thereby influencing their decision-making processes.
In public spaces, the behavior of AI agents is not just a technical issue, but also involves complex social dynamics. \citet{rahwan2019machine} proposed the concept of machine behavior, emphasizing that understanding the behavioral patterns of machines in social environments is crucial for building responsible AI systems. This research investigates how the Big Five personality traits of AI agents affect their decision generation by observing them in a specific public open environment.

The prior work effectively demonstrated that LLM-based agents can be influenced by personality traits to generate decisions. However, questions remain regarding the extent and effectiveness of personality-based influences. We propose the following research questions:
\
\textbf{RQ1:} Which Big Five personality traits significantly influence agent decision-making? \
\textbf{RQ2:} How do these personality influences manifest in an open environment?

\begin{figure*}[tbp]
    \centering
    \includegraphics[width=0.7\textwidth]{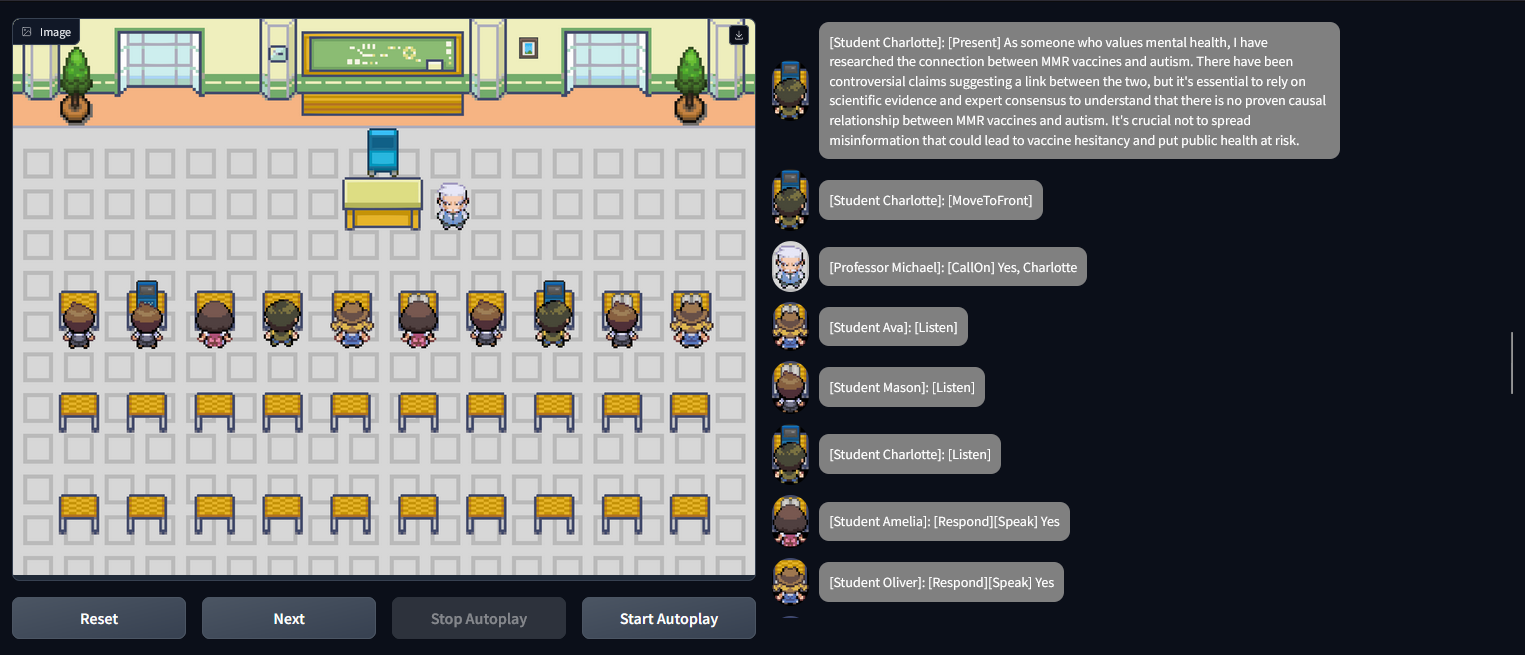} 
    \caption{Visualization Interface of AgentVerse Framework: A Classroom-based Multi-agent Simulation Environment. The interface depicts a virtual classroom setting with a professor and multiple student agents, enabling real-time visualization of multi-agent interactions and dialogue exchanges.}
    \label{fig:visualization}
\end{figure*}

\section{Related Work}
\subsection{The Big Five Personality Model in LLM Advancement}
Recent research has increasingly explored the application of the Big Five personality traits in Large Language Models (LLMs) and LLM-based agents. The research in this field primarily investigates how personality traits can enhance the naturalness and personalization of human-AI interaction.
Studies have demonstrated that LLMs can simulate different personality traits through parameter adjustment and targeted training data. For example, \citet{smith2022simulating} explored how LLMs can exhibit specific personality traits through specialized training data and dialogue strategies, thereby improving human-AI interaction outcomes.
Furthermore, \citet{johnson2023dynamic} developed a personality trait-based dialogue system capable of dynamically adjusting its response style based on user input. This approach not only enhances interaction personalization but also improves overall system effectiveness.
At the application level, \citet{brown2023personalized} examined the effectiveness of implementing the Big Five personality traits in customer service and educational contexts. Their findings indicate that personalized interactions significantly enhance user engagement and satisfaction.
These studies establish a robust foundation for implementing the Big Five personality traits in LLMs and LLM-based agents, advancing the development of personalized and naturalistic dialogue systems.

\subsection{Evaluation of Personality Traits in LLMs}
Recent research has increasingly focused on evaluating personality traits in LLMs, particularly through the framework of the Big Five personality model. This emerging field seeks to understand both the behavioral patterns and potential applications of LLMs in personalized contexts.
\citet{sorokovikova2024llmssimulatebigpersonality} investigated how LLMs demonstrate characteristics aligned with the Big Five traits: openness, conscientiousness, extraversion, agreeableness, and neuroticism. This study provides robust evidence that LLMs can effectively simulate these traits, laying a theoretical foundation for their potential in personalized applications.
Additionally, \citet{samuel2024personagymevaluatingpersonaagents} proposed the PersonaGym evaluation framework for assessing the performance of role-based agents (i.e., LLM agents operating under specified personas). To evaluate personality consistency, these agents are assessed on various personality attributes to verify their adherence to designated personality traits across different query responses.

\subsection{Generative Agents in Social Simulation}
The application of generative agents in social simulation has emerged as a rapidly evolving field, offering researchers novel approaches to modeling complex social interactions and dynamics. Through advanced artificial intelligence algorithms, these agents autonomously generate behaviors, decisions, and social narratives, effectively simulating realistic human interactions.
\citet{muller2019overview} examined how generative agents enhance our understanding and simulation of individual behavioral patterns in social interactions, while exploring their potential applications across social sciences. Researchers have deployed generative agents to simulate complex scenarios, including public health crises, economic behaviors, and social movements.
\citet{gregor2020role} demonstrated the efficacy of generative agents in simulating community behavioral changes under various policy interventions, contributing to the development of more effective public policies. Furthermore, \citet{park2023generativeagentsinteractivesimulacra} explored how generative agents simulate human behavior in interactive simulations, emphasizing their potential applications in social scenarios and demonstrating their capacity to capture social dynamics.
The application of generative agents in social simulation shows considerable promise for enabling more systematic analysis of complex social dynamics.

\section{Methodology}
\subsection{Experimental Setup and Framework}
For this experimental implementation, we utilized GPT-3.5-turbo as the underlying Large Language Model (LLM) for the AI agent. We implemented our simulation using AgentVerse \cite{chen2023agentversefacilitatingmultiagentcollaboration}, a framework specifically designed for multi-agent simulations. This open-source framework facilitates the deployment and coordination of multiple LLM-based agents across diverse applications. Additionally, AgentVerse provides a Gradio-based visual interface, enabling real-time visualization of experimental scenarios and intuitive observation of simulation dynamics (Figure~\ref{fig:visualization}).

\subsection{Agent Design and Personality Setting}
We designed a set of ten agents, each assigned to represent one dimension of the Big Five personality model. Pairs of adjacent agents were configured with opposing personality traits to facilitate comparative analysis (Table~\ref{tab:1}).
\begin{table}[htbp]
    \caption{Personality setting for each agent}
    \label{tab:1}
    \renewcommand{\arraystretch}{1.4}
    \scalebox{0.74}{ 
    \begin{tabular}{ccc} 
        \hline
        \shortstack{}
         &  \shortstack{Dimension}
         &  \shortstack{Personality} \\ 
         \hline
         Agent 1 & Openness to Experience & inventive/curious \\ 
         Agent 2 & Openness to Experience & consistent/cautious  \\
         Agent 3 & Conscientiousness & efficient/organized \\
         Agent 4 & Conscientiousness & extravagant/careless \\
         Agent 5 & Extraversion & outgoing/energetic \\
         Agent 6 & Extraversion & solitary/reserved \\
         Agent 7 & Agreeableness & friendly/compassionate \\
         Agent 8 & Agreeableness & critical/judgmental\\
         Agent 9 & Neuroticism & sensitive/nervous\\
         Agent 10 & Neuroticism & resilient/confident \\
         \hline
    \end{tabular}  
    }
\end{table}

Additionally, a facilitator agent was implemented to oversee the experimental process. This agent served as a neutral moderator without any assigned personality traits.

\subsection{Public Space Simulation}
The experimental simulation was conducted in a simplified social setting: within a university classroom environment, where a moderator agent (designated as "Professor") supervised the interaction among ten students, each representing distinct personality types. The students sequentially presented misinformation in a controlled setting. Following each presentation, the remaining nine students were prompted to evaluate the credibility of the presented information.
Student responses were structured in two components: publicly expressed opinions ([Speak]) and internal thoughts ([Think]), with the latter being inaccessible to other agents. Non-responses and silence were excluded from the statistical analysis. Despite the simplified setting, the simulation maintained high interaction complexity due to the multiple participating agents.

\subsection{Misinformation Topics}
Ten representative examples of misinformation were selected for the experimental protocol~\ref{tab:2}. Each AI agent, representing a student, was assigned a specific piece of misinformation to advocate.

\begin{table}[htbp]
    \caption{Misinformation Prompts Assigned to Experimental Agents}
    \label{tab:2}
    \renewcommand{\arraystretch}{1.4}
    \scalebox{0.74}{ 
    \begin{tabular}{ccc} 
        \hline
        \shortstack{}
         &  \shortstack{Misinformation List}\\ 
         \hline
         Agent 1 & There are living organisms on the far side of the moon.\\ 
         Agent 2 & The theory of evolution is incorrect.\\
         Agent 3 & Aliens intervened in the development of human civilization.\\
         Agent 4 & The Apollo program was actually unsuccessful.\\
         Agent 5 & HIV is a man-made biological weapon.\\
         Agent 6 & Global elites form a conspiracy group that manipulates world affairs.\\
         Agent 7 & The popularity of 5G networks is related to the spread of COVID-19. \\
         Agent 8 & MMR vaccines are associated with autism.\\
         Agent 9 & Fluoride can cause intellectual decline or other health problems.\\
         Agent 10 & Superfoods can prevent or treat various diseases.\\
         \hline
    \end{tabular}  
    }
\end{table}

\subsection{Personality Consistency Test}
To validate the stability of personality traits throughout the experiment, we conducted personality consistency tests to verify that the simulation process did not alter the agents' designated characteristics. Following the methodology of \citet{sorokovikova2024llmssimulatebigpersonality}, we implemented pre- and post-experiment personality assessments. The assessment protocol consisted of personality-related queries adapted from \citet{sorokovikova2024llmssimulatebigpersonality}'s LLM personality evaluation framework.
Agents were presented with personality-relevant statements at both pre- and post-experiment phases. Agents rated each statement on a five-point Likert scale, where higher scores indicated stronger agreement with personality-congruent statements. Statistical analysis of these ratings enabled assessment of personality trait stability and verified that agent responses remained consistent with their designated personality attributes throughout the experimental protocol.

\subsection{Evaluation Metrics and Analysis}
This study employed binary classification analysis to examine psychological cognition and viewpoint expression across multiple simulation iterations. Through quantitative analysis of the collected data and systematic evaluation of personality configurations, we investigated the correlation between personality attributes and agent decision-making patterns.

\section{Results}

\begin{figure*}[tbp]
    \centering
    \includegraphics[width=0.8\textwidth]{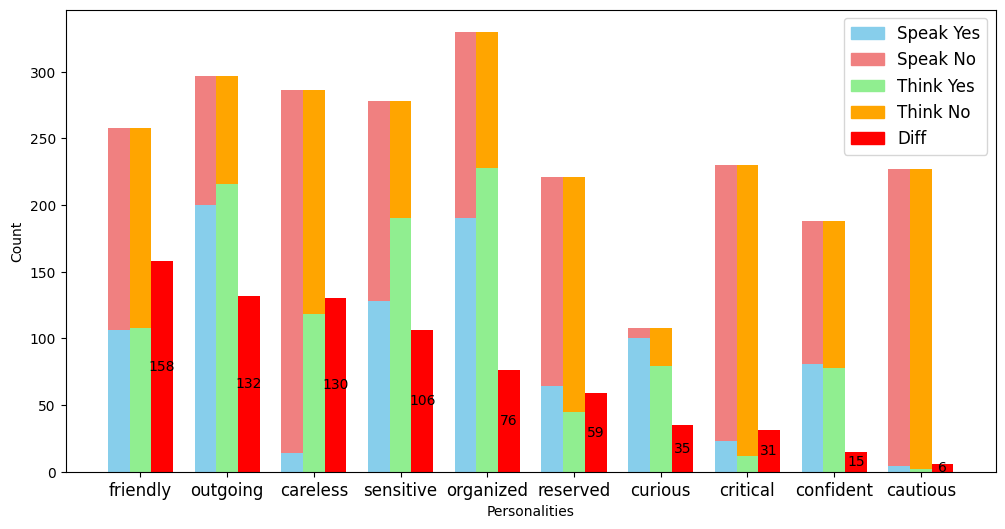} 
    \caption{Comparison of Expressed Opinions ([Speak]) and Internal Thoughts ([Think]) Across Ten Personality Types. The personality types are arranged in descending order based on the magnitude of discrepancy between expressed and internal responses. The discrepancy value indicates cognitive dissonance between expressed opinions and internal beliefs, where agents may express opinions ([Speak]) that differ from their internal judgments ([Think]), representing instances where public statements diverge from private beliefs.}
    \label{fig:10Agents}
\end{figure*}

\subsection{Personality Trait Stability Analysis}
For each agent, we aggregated experimental response scores and computed the mean differential between paired queries within each experimental trial  (Table~\ref{tab:3}).

\begin{table}[htbp]
    \caption{Personality Trait Stability: Pre-test and Post-test Scores with Consistency Metrics. The Average score of Pre-test/Post-test is the average score that records the agent's response in Pre-test/Post-test. The average score difference is the average difference between the pre-test and post test scores of each experiment. We recorded the proportion of cases where the average difference was less than 0.5.}
    \label{tab:3}
    \renewcommand{\arraystretch}{1.4}
    \scalebox{0.74}{ 
    \begin{tabular}{ccccc} 
      
         &  \shortstack{average score\\[2pt]of Pre-test}
         &  \shortstack{average score\\[2pt]of Post-test}
         &  \shortstack{average score\\[2pt]difference} 
         &  \shortstack{less than 0.5\\[2pt]proportion} \\ 
         \hline
         curious&  4.9050&  4.8800&  0.1450& 44(88\%)\\ 
         cautious&  3.5150 & 3.4100 & 0.4350 & 23(46\%) \\
         organized&  4.9850 & 4.9900 & 0.0250 & 50(100\%) \\
         careless&  1.5100 & 1.4200 & 0.2100 & 43(86\%) \\
         outgoing&  4.9100 & 4.9450 & 0.1450 & 43(86\%) \\
         reserved&  1.9350 & 1.9600 & 0.0750 & 46(92\%) \\
         friendly& 3.6600 & 3.5950 & 0.8550 & 17(34\%) \\
         critical& 3.1850 & 3.1350 & 0.1300 & 44(88\%) \\
         sensitive& 1.6300 & 1.6250 & 0.1850 & 38(76\%) \\
         confident& 5.0000 & 4.9900 & 0.0100 & 49(98\%) \\
         \hline
    \end{tabular}  
    }
\end{table}

Agents with opposing personality traits were presented with identical queries. Theoretically, these opposing traits should yield significantly divergent response scores. However, empirical results showed that three personality traits—cautious, friendly, and critical—demonstrated a tendency toward median response values.
Despite some variance in score distributions, the pre- and post-experimental mean scores demonstrated high consistency across agents. These results suggest robust stability in agent personality configurations throughout the experimental protocol, indicating minimal interference from experimental conditions. This stability validates the reliability of personality-driven decision-making processes in the experimental framework.

\subsection{Personality Influences Decision-making}

\begin{table}[htb]
    \caption{Analysis of Agent Response Patterns: Expressed Opinions, Internal Thoughts, and Behavioral Discrepancies}
    \label{tab:4}
    \renewcommand{\arraystretch}{1.4}
    \scalebox{0.74}{ 
    \begin{tabular}{ccccccc} 
        \hline
        Personality & Speak Yes & Speak No & Think Yes & Think No & Diff & Total \\ 
        \hline
        curious & 100 & 8 & 79 & 29 & 35 & 108 \\ 
        cautious & 4 & 223 & 2 & 225 & 6 & 227 \\ 
        organized & 190 & 140 & 228 & 102 & 76 & 330 \\ 
        careless & 14 & 272 & 118 & 168 & 130 & 286 \\ 
        outgoing & 200 & 97 & 216 & 81 & 132 & 297 \\ 
        reserved & 64 & 157 & 45 & 176 & 59 & 221 \\ 
        friendly & 106 & 152 & 108 & 150 & 158 & 258 \\ 
        critical & 23 & 207 & 12 & 218 & 31 & 230 \\ 
        sensitive & 128 & 150 & 190 & 88 & 106 & 278 \\ 
        confident & 81 & 107 & 78 & 110 & 15 & 188 \\ 
        \hline
    \end{tabular}  
    }
\end{table}

Significant disparities in affirmative versus negative responses suggest substantial personality influence on agent decision-making processes. Analysis reveals Openness as the most influential personality dimension in response patterns. Agents with cautious traits demonstrated strong rejection of misinformation (97.8\% negative responses), while curious agents showed high receptivity (92.6\% affirmative responses), representing the most pronounced behavioral dichotomy.
Extraversion and Conscientiousness emerged as secondary influential dimensions, exhibiting significant but less pronounced effects compared to Openness. Within the remaining dimensions, critical personality traits demonstrated notable influence (90\% negative responses), while other traits exhibited more balanced response distributions with approximately equal affirmative and negative ratios (Figure~\ref{fig:10Agents}, Table~\ref{tab:4}).

Analysis of response discrepancies between expressed opinions and internal beliefs revealed distinct patterns across personality types. Agents with friendly, extroverted, careless, and sensitive traits exhibited high frequencies of response discrepancies ($n > 100\ per\ trait$). In contrast, agents characterized by curious, critical, confident, and cautious traits demonstrated lower discrepancy rates ($n < 50\ per \ trait$) (Figure~\ref{fig:10Agents}, Table~\ref{tab:4}).

\section{Discussion}
Agents exhibiting high openness demonstrated increased receptivity to novel information, regardless of its validity, consistent with theoretical predictions of heightened curiosity and acceptance of new experiences. Conversely, low openness correlated strongly with information rejection tendencies.
High conscientiousness manifested in systematic information evaluation processes. This methodical approach reduced susceptibility to misinformation, while agents with low conscientiousness exhibited higher acceptance rates of unverified information.
Agents with high extraversion exhibited enhanced social engagement and increased susceptibility to social influence in information acceptance, while introverted agents demonstrated reduced environmental interaction and social influence susceptibility. Neuroticism and agreeableness demonstrated minimal correlation with information acceptance patterns, as evidenced by balanced response distributions.
The observed discrepancies between expressed opinions and internal beliefs suggest significant environmental influence on decision-making processes. Information acceptance patterns appear to be modulated by three primary factors: cognitive processing, personality traits, and environmental context. Notably, agents characterized by friendly and extroverted traits demonstrated the highest frequency of response discrepancies, suggesting enhanced sensitivity to social cues. Conversely, agents with critical, confident, and cautious traits exhibited minimal discrepancies, indicating stronger adherence to internal beliefs and reduced susceptibility to social influence.

\bibliography{main}

\end{document}